# Enhanced sinterability and *in vitro* bioactivity of diopside through fluoride doping

E. Salahinejad *, M. Jafari Baghjeghaz

Faculty of Materials Science and Engineering, K.N. Toosi University of Technology, Tehran, Iran

**Abstract**

In this work, diopside ($CaMgSi_2O_6$) was doped with fluoride at a level of 1 mol.%, without the formation of any second phase, by a wet chemical precipitation method. The sintered structure of the synthesized nanopowders was studied by X-ray diffraction, Fourier transform infrared spectroscopy and field-emission scanning electron microscopy. Also, the samples' in vitro apatite-forming ability in a simulated body fluid was comparatively evaluated by electron microscopy, inductively coupled plasma spectroscopy and Fourier transform infrared spectroscopy. According to the results, the material's sinterability was improved by fluoride doping, as realized from the further development of sintering necks. It was also found that compared to the undoped bioceramic, a higher amount of apatite was deposited on the surface of the doped sample. It is concluded that fluoride can be considered as a doping agent in magnesium-containing silicates to improve biological, particularly bioactivity, behaviors.

* Corresponding Author:
Email Addresses: <salahinejad@kntu.ac.ir>, <erfan.salahinejad@gmail.com>

## 1. Introduction

Diopside ($CaMgSi_2O_6$) with a monoclinic crystal structure and density of on average 3.4 gr/cm3 belongs to the group of magnesium-containing silicate ceramics. Due to suitable mechanical properties, biocompatibility and osteoinduction, diopside has been used in a wide range of biomedical applications like: coating on biomedical implants [1-3], bone and tooth tissue engineering [4, 5], drug delivery [6], in vivo imaging [7] and surgery hemostasis applications [8]. Concerning bioactivity, although diopside can more or less offer *in vitro* apatite formation and *in vivo* bone formation, this characteristic is not perfect and should be directed for its further development.

Doping of proper species can be an efficient approach to improving bioactivity, mechanical properties and anti-bacterial properties. One of the elements that is frequently used for doping in apatite ceramics and glasses is fluorine. Fluoride in bioceramics, on the one hand, can inhibit the demineralization of the enamel and dentin and bacterial enzyme activity, and on the other hand, can enhance remineralization and bioactivity [9]. In addition, this type of dopants improves the chemical stability and mechanical properties of hydroxyapatite [10, 11]. Typically, it has been shown that the addition of fluoride results in the formation of fluorapatite which is more acid-resistant than carbonated hydroxyapatite [12], making it interesting for dentistry applications.

To the best of our knowledge, doping with fluoride has not been considered in magnesium-containing biosilicates, including in diopside. Thus, in this study, diopside was originally doped with 1 mol.% fluoride by a wet chemical (coprecipitation) method using chloride inorganic compounds. Chloride precursors have been previously used for the chemical synthesis of other multi-component ceramics [13-23]. Afterwards, the resultant structure (including phase formation, bonding and sintering evolution) and *in vitro* apatite-forming ability in a simulated body fluid were investigated.

## 2. Experimental procedure

*2.1. Materials*

To synthesize diopside, calcium chloride ($CaCl_2$, Merck, >98%), magnesium chloride ($MgCl_2$, Merck, >98%) and silicon tetrachloride ($SiCl_4$, Merck, >99%) as the fundamental precursors of Ca, Mg and Si, respectively, and magnesium fluoride ($MgF_2$, Alfa Aesar, >99%) as the source of F were used in this work. Also, dry ethanol ($C_2H_5OH$, Merck, >99%) and aqueous ammonia solution ($NH_4OH$, Merck, 25%) were employed as the solvent and precipitating agent, respectively.

*2.2. Precipitation synthesis*

For undoped diopside synthesis, an equimolar amount of $CaCl_2$ and $MgCl_2$ was dissolved in ethanol. After complete dissolution, the proper content of $SiCl_4$ (according to the stoichiometry) was added to the above solution while stirred for 30 min in an ice-water bath. A proper content of the ammonia solution was added to the above solution under magnetic stirring to reach pH = 10. The synthesis of F-doped diopside was similar to the above procedure, with this difference that $MgF_2$ (at an mount to incorporate 1 mol.% of fluoride in diopside) was added shortly after the addition of $SiCl_4$ and before adding the ammonia solution and precipitation. In addition, to keep the molar ratio of Ca:Mg:Si = 1:1:2 according to the diopside stoichiometry ($CaMgSi_2O_6$), the additional molar value of Mg caused by $MgF_2$ was subtracted from $MgCl_2$ loaded at the beginning of the process. Lastly, the obtained white precipitates were dried at 120 °C for 6 h in an oven, and then analyzed by X-ray diffraction (XRD). Also, transmission electron microscopy (TEM) was used to observe the synthesized precipitates after washing and centrifuging with water to remove coproducts of the precipitation process.

*2.3. Sintering*

The dried powders were ground in an agate mortar, uniaxially cold-pressed at 100 MPa, and finally sintered at 1200 °C for 2 h in air at a heating and cooling rate of 10 °C/min. The sintered samples were characterized by XRD for phase identification, Fourier transform infrared spectroscopy (FTIR) for bonding identification and F-doping confirmation, and Field-emission scanning electron microscopy (FESEM) for observing sintering necks between the particles as a sign of sintering progress.

*2.4. Apatite-forming ability*

To realize the effect of F-doping on the bioactivity of diopside, the sintered samples were immersed in the simulated body fluid (SBF) [24] for 3 days at 37 °C. Afterwards, the dried samples were studied by FESEM equipped with energy-dispersive X-ray spectroscopy (EDS) and FTIR. To further analyze apatite-forming ability, *in vitro* biodegradation was also evaluated by studying the concentration of principal ions in the SBF before and after immersion by inductively coupled plasma spectroscopy (ICP). The pH value of the SBF before and after immersion was also measured by a Sartorius Professional Meter PP-15.

**3. Results and discussion**

The TEM micrograph of the coprecipitation product after washing is indicated in Fig. 1. Washing was performed to remove the coproduct of the coprecipitation process, which will be below shown to be ammonium chloride based on XRD, and to merely consider the main silicate-based particles. The polygonal particle appearing a different contrast compared to the other particles may be a residual ammonium chloride particle. It can be seen that the majority of the powder particles are approximately 15 nm in size with a relatively faceted morphology.

Fig. 2(a) shows the XRD pattern of the powder dried at 120 °C. As can be seen, the pattern includes the characteristic peaks of ammonium chloride ($NH_4Cl$, Ref. code: 01-073-0365) and ammonium aqua magnesium chloride ($NH_4(Mg(H_2O)_6)Cl_3$, Ref. code: 01-076-1454). Considering the loaded precursors, it can be inferred that the dried powder has an amorphous phase containing the Ca and Si species. According to Fig. 2(b), the sample sintered at 1200 °C is merely composed of diopside ($MgCaSi_2O_6$, Ref. code: 00-017-0318). Compared to the sample dried at 120 °C, during heating to 1200 °C, on the one hand, ammonium chloride goes away from the material via sublimation; on the other hand, the amorphous phase and ammonium aqua magnesium chloride react to form pure diopside and some volatile byproducts. According to the XRD analysis, as demonstrated in Fig. 2(c), the sample doped with 1 mol.% of fluoride also consists of pure diopside without the formation of any second phase. Also, a precise comparison between Figs. 2(b) and 1(c) suggests that F-doping enhances the sharpness and intensity of the diffraction peaks which are directly related to crystallinity and crystallite size. The crystallite sizes of diopside for the sintered, undoped and doped samples, calculated by the Scherrer equation from the full width at half maximum values of the most intense diffraction peak (2θ = 29.9 °), were 52 and 66 nm, respectively. It can be due to the fact that fluoride lowers the melting point of silicates [25];

thus, the sintering temperature of 1200 °C yields a higher homologous temperature for the F-doped sample. The higher homologous temperature can increase crystallinity and crystallite size at a given firing temperature, explaining the difference detected in the XRD peaks of the sintered samples.

In order to further characterize the structure of the synthesized materials and to verify the incorporation of F into diopside, the sintered samples were analyzed by FTIR. For the pure diopside, based on Fig. 3(a), the non-bridging bending vibrations of O-Ca-O and O-Mg-O lie at around 400 cm$^{-1}$ and in the range of 465 cm$^{-1}$ to 525 cm$^{-1}$, respectively. Also, sharp dual peaks at almost 635 and 672 cm$^{-1}$ are assigned to the O-Si-O bending mode, whereas broad peaks at around 860, 960, 1070 cm$^{-1}$ correspond to the Si-O symmetric stretching mode in the SiO$_4$ tetrahedron. That is, the FTIR analysis well exhibits all of the typical functional groups related to diopside, which is also in agreement with Refs. [26, 27]. However, for the F-doped silicate, as well as the functional groups of diopside, as indicated in Fig. 3(b), additional adsorption bands appear at 800 and 930 cm$^{-1}$ which correspond to the Si-F stretching mode [28-30]. Also, as shown in Fig. 3, the majority of the Si-O vibrations detected for pure diopside slightly shifted to higher wavenumbers for 2-5 cm$^{-1}$ as a result of F-doping. Fluoride is partially substituted for oxygen, albeit the population of the Si-O bonds is still more than the Si-F bonds, considering the doped amount. Since the electronegativity of F is higher than O and the Si-F bond is essentially stronger than the remaining Si-O bond; the charge distribution on the Si-O bonds is changed in term of dragging electrons from O. It gives rise to an increase in the covalence of the Si-O bonds and thereby an increase its IR vibration frequencies, which is compatible with Ref. [31]. The typical decrease of the 635 and 672 cm$^{-1}$ absorption bands of the O-Si-O bending is another evidence for the partial replacement of F for O, since the FTIR peak intensity is directly proportional to the population of the related bond. In conclusion, the FTIR analysis confirmed the incorporation of fluoride into diopside via substitution for oxygen.

Fig. 4 indicates the SEM micrograph of the sintered samples before immersion in the SBF, suggesting three typical points. First, both the samples show a porous structure mainly originating from the sublimation of ammonium chloride produced during coprecipitation. Theoretically, the dried sample contains 12 mole of $NH_4Cl$ per each mole of diopside, taking account of the stoichiometric amount of Cl in the precursors. Considering their density, it can be estimated that the volume ratio of $NH_4Cl$ to diopside is almost 6.5 in the dried powder. This significant content of $NH_4Cl$ sublimes during heating with no sign of spoiling and leaves a considerable level of interconnecting open pores in the sintered samples, which is adventurous for bioactivity via increasing the surface area. Second, the average particle size for the doped sample is larger than that for the undoped silicate, where it is 300 and 800 nm in the undoped and doped samples, respectively. Third, the mean diameter of sintering necks in F-doped diopside is about 500 nm, while this parameter is also almost 150 nm for the undoped ceramic, which is indicative of an improved sintering evolution caused by F-doping. The larger particle size and improved sinterability for the doped particles are due to the higher homologous temperature experienced by this sample during sintering, as pointed out above.

The FESEM micrographs of the sintered samples after 3 days of immersion in the SBF is presented in Fig. 5. An overall scan on the low-magnification micrographs shows that almost no/little precipitate is formed on the surface of the undoped sample, as represented in Fig. 5(a). Albeit, in high-magnification micrograph shown in Fig. 5(b), some nanometric precipitates can be discovered on a few of the undoped diopside areas. As indicated in Fig. 5(c), the EDS analysis on these precipitates confirms the presence of phosphorous, suggesting that they can be a kind of apatite. Since the Mg and Si peaks are also detected in the EDS analysis and confidently related to the diopside matrix, the Ca trace belong to both the matrix and precipitates. Thus, the ratio of Ca/P cannot be used to identify the type of the apatite phase formed in this work. In contrast to the undoped sample, almost 85 % of the doped diopside surface is covered by uniform leaf-like apatite precipitates (Fig. 5(d)), where the mean thickness of the apatite plates is 20 nm based on Fig. 5(e). Again, the EDS analysis (Fig. 5(f)) infers that the precipitates are apatite. The higher relative intensity of P for the doped sample, in comparison to the undoped ceramic, is another indicative of the better evolution of apatite on the doped one.

The FTIR spectra of the samples after soaking in the SBF are shown in Fig. 6. It is known that there are several overlaps between the FTIR absorption bands of the silicate group (including in diopside), on the one hand, and the phosphate [27, 32], hydrogen phosphate [33] and carbonate groups [34], on the other hand. This, therefore, challenges the straightforward distinction of silicates and apatites through FTIR. The typical overlaps are associated to the bending (480 and 500 $cm^{-1}$) and stretching modes (970 and 1080 $cm^{-1}$) of phosphate, the vibration modes of hydrogen phosphate (875 and 960 $cm^{-1}$), and the bending mode of carbonate (875 $cm^{-1}$). However, a comparison between Figs. 3 and 6 for both the samples signifies that the intensity of the aforementioned overlapping IR bands increases after soaking in the SBF. It suggests the presence of the phosphate and carbonate functional groups in the precipitates. Also, the new small peak at about 1000 $cm^{-1}$ is assigned to a vibration mode of phosphate for the doped sample. The carbonate functional group is detected from the vibration bands located at about 1420, 1460 and 1550 $cm^{-1}$. On the other hand, in the undoped sample (Fig. 6(a)), the low intensity of the vibration mode of hydroxyl (3570 $cm^{-1}$) in apatite suggests that the carbonate group partially substitutes for hydroxyl [35]. However, in the doped sample during soaking in the SBF, the hydroxyl group of apatite seems to be completely substituted by carbonate as well as fluoride [36, 37], since the vibration mode at almost 3570 $cm^{-1}$ disappears (Fig. 6(b)). In conclusion, the FTIR analysis on the samples immersed in the SBF signifies the deposition of hydroxycarbonate/fluorocarbonate apatite which is a favorable phase to induce the chemical or bioactive fixation of bioceramics to natural tissues. Also, it can be seen that the sharpness of the apatite peaks for the doped sample is more than the pure diopside, verifying the higher content of apatite deposited on the doped sample, as realized from the FESEM studies (Fig. 5).

The study of the samples' *in vitro* biodegradation via evaluating the Ca, Mg, Si and P concentrations in the SBF before and after immersion, as conducted by ICP (Fig. 7), can be correlated to the apatite-forming ability. Before immersion, Ca and M exist in both the SBF and diopside; nevertheless, Si only exist in diopside and P only lies in the SBF. After immersion, the Ca content in the SBF is determined by compromising between the precipitation of apatite and the dissolution of diopside and apatite formed, which reduces and enhances the Ca content of the SBF, respectively. Thus, the amount of Ca in the SBF after immersion can be not surely regarded as a criterion to compare the apatite-forming ability of the immersed samples. This is in agreement with the literature [38-41] where although the apatite formation on the surface of Ca-containing bioceramics is developed, the Ca content in the SBF can be first increased and then decreased by increasing immersion time. It is, however, noticeable that the higher amounts of Ca for both the sample after immersion compared to those before immersion, due to the matrix dissolution, is critical for the apatite formation via providing a supersaturated solution. The enhancement of the Si and Mg concentration due to immersion is attributed to their dissolution from the samples into the SBF, where the increases for the doped sample are lower than the pure sample. The more limited dissolution of Si and Mg as a result of F-doping can be due to the higher electronegativity of F than O which provides stronger bonding of these two ions to the ceramic network (i.e. Mg-F and Si-F bonds are stronger than Mg-O and Si-O ones, respectively). Since these two ions are not incorporated into the apatite precipitated, monitoring their contents in the SBF provides no clear guidance to compare the *in vitro* bioactivity of the samples. Contrary to the previous ions, the only source of changes in the P content is the precipitation of apatite. Accordingly, the less P content in the SBF means the more apatite formation on the ceramic surface. Because of the considerably lower amount of P for the SBF soaking the doped sample realized from the ICP analysis, it can be concluded that the apatite-forming ability of diopside is improved by 1 mol.% F-doping, confirming the FESEM studies (Fig. 5).

The pH variations of the SBF as a result of immersion of the samples can be correlated to the apatite-forming ability. Before immersion, the SBF had pH = 7.4, while the pH value reached 8.3 and 7.6 after 3 days of immersion of the undoped and doped samples, respectively. According to the ICP tests, diopside releases $Ca^{2+}$ and $Mg^{2+}$ into the surrounding SBF, thereby leaving cation vacancies near the surface. These vacancies are substituted for $H^+$ ions of the solution, which leads to an increase in pH from 7.4 to 8.3. But for the F-doped diopside, $F^-$ ions released is exchanged for $OH^-$ ions of the SBF and tends to

This is the accepted manuscript (postprint) of the following article:
M. Jafari Baghjeghaz, E. Salahinejad, *Enhanced sinterability and in vitro bioactivity of diopside through fluoride doping*, Ceramics International, 43 (2017) 4680-4686.
https://doi.org/10.1016/j.ceramint.2016.12.144decrease pH. It means that the release of $F^-$ ions slightly buffers the effect of the $Ca^{2+}$ and $Mg^{2+}$ release and decreases pH from 8.3 to 7.6, which is in agreement with Refs. [12, 42].

The different apatite-forming ability of the sintered samples as a result of F-doping into diopside can be explained as follows. On the one hand, as shown in Fig. 4, F-doping increased the particle size, while it is known that smaller particle sizes via providing larger surface area are advantageous for bioactivity. On the other hand, the SEM observations (Fig. 5) and the consideration of the phosphorous content (Fig. 7) suggested that the bioactivity of diopside was improved by F-doping. Thus, it is concluded in this study, the contribution of the chemical composition prevails over that of the particle size in determining apatite-forming ability. Concerning why and how fluoride improves bioactivity, in the literature [10-12, 40, 42], it is attributed to the formation of fluorapatite instead of carbonated hydroxyapatite, where that former is more stable than the latter. However, the F content used in the current work is so low that it cannot lead to the deposition of fluorapatite. For this case, the below reasons can be involved:

*i*) As indicated in the FTIR analysis (Fig. 6), a part of the hydroxyl group of the apatite-like precipitates was substituted by fluoride for the doped sample. This increases the chemical stability of apatite and retards its re-dissolution, inducing better *in vitro* and *in vivo* bioactivity [43].

*ii*) According to the pH measurements presented above, as a result of F-doping into diopside, the pH value of the SBF after immersion was reduced from 8.3 to 7.6, which was attributed to the exchange of $F^-$ and $OH^-$. That is, the concentration of $OH^-$ on the ceramic surface is enhanced by F-doping, developing silanol (Si–OH) groups on the surface. These silanol (Si–OH) groups play a significant role in the apatite nucleation [44]. Thus, the addition of fluoride to diopside via providing more silanol groups on the surface accelerates the apatite nucleation and leads to a better *in vitro* bioactivity.

**4. Conclusions**

In this work, 1 mol.% of fluoride was added to diopside via a coprecipitation method; and the resultant structure and bioactivity were studied. According to the results, the incorporation of F into diopside was confirmed without the formation of any second phase. Also, sintering necks between the synthesize nanoparticles were developed by adding fluoride into diopside, which is indicative of improved sinterability. The apatite-forming ability of diopside was improved by F-doping, as correlated with the pH value and ionic concentration of the SBF after soaking the samples. It is eventually concluded that fluoride can be considered as a doping agent in magnesium-containing silicates to improve biological, particularly bioactivity, behaviors.

**This is the accepted manuscript (postprint) of the following article:**
M. Jafari Baghjeghaz, E. Salahinejad, *Enhanced sinterability and in vitro bioactivity of diopside through fluoride doping*, Ceramics International, 43 (2017) 4680-4686.
https://doi.org/10.1016/j.ceramint.2016.12.144
[42] D.S. Brauer, N. Karpukhina, D. Seah, R.V. Law, R.G. Hill, Fluoride-containing bioactive glasses, Advanced Materials Research, Trans Tech Publ, 2008, pp. 299-304.

[43] P. Taddei, E. Modena, A. Tinti, F. Siboni, C. Prati, M.G. Gandolfi, Effect of the fluoride content on the bioactivity of calcium silicate-based endodontic cements, Ceramics International, 40 (2014) 4095-4107.

[44] C. Ohtsuki, T. Kokubo, T. Yamamuro, Mechanism of apatite formation on CaO SiO 2 P 2 O 5 glasses in a simulated body fluid, Journal of Non-Crystalline Solids, 143 (1992) 84-92.
15

**Figures**

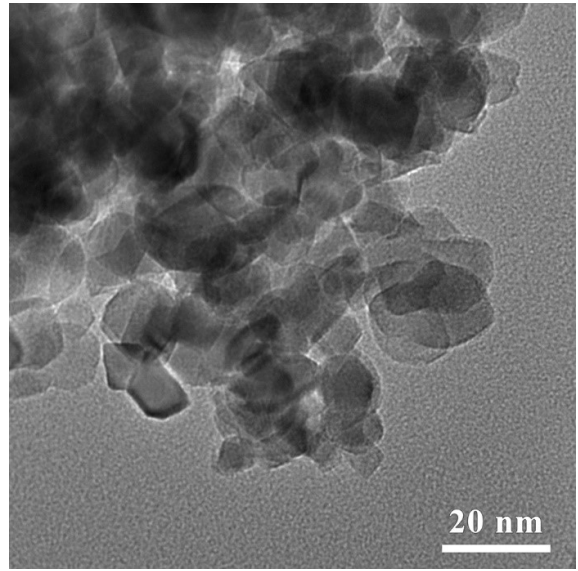

Fig. 1. TEM micrograph of the washed powder sample.

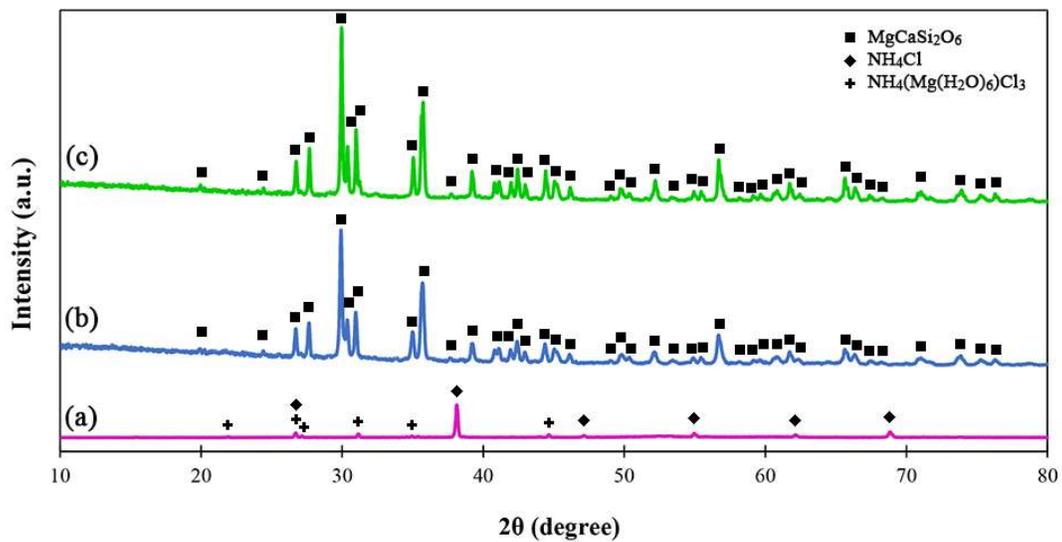

Fig. 2. XRD patterns of the sample dried at 120 °C (a) and sintered at 1200 °C at the pure state (b) and with the F dopant addition (c).

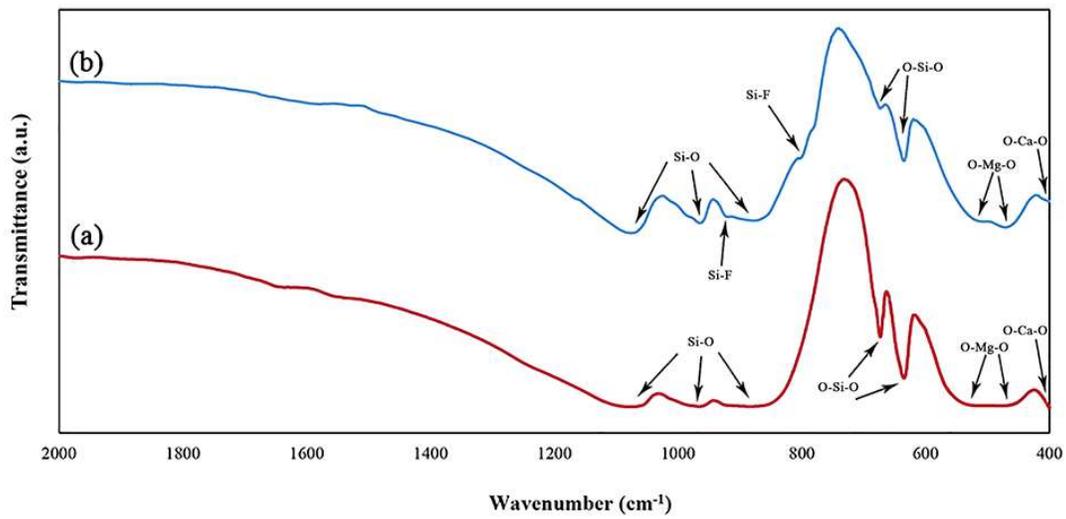

Fig. 3. FTIR spectra of the sintered, undoped (a) and doped (b) samples before immersion in the SBF.

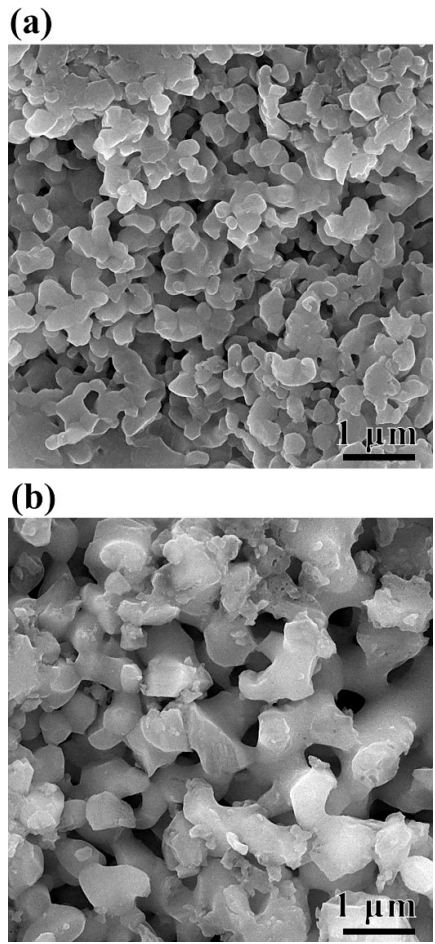

Fig. 4. SEM micrograph of the sintered, undoped (a) and doped (b) samples.

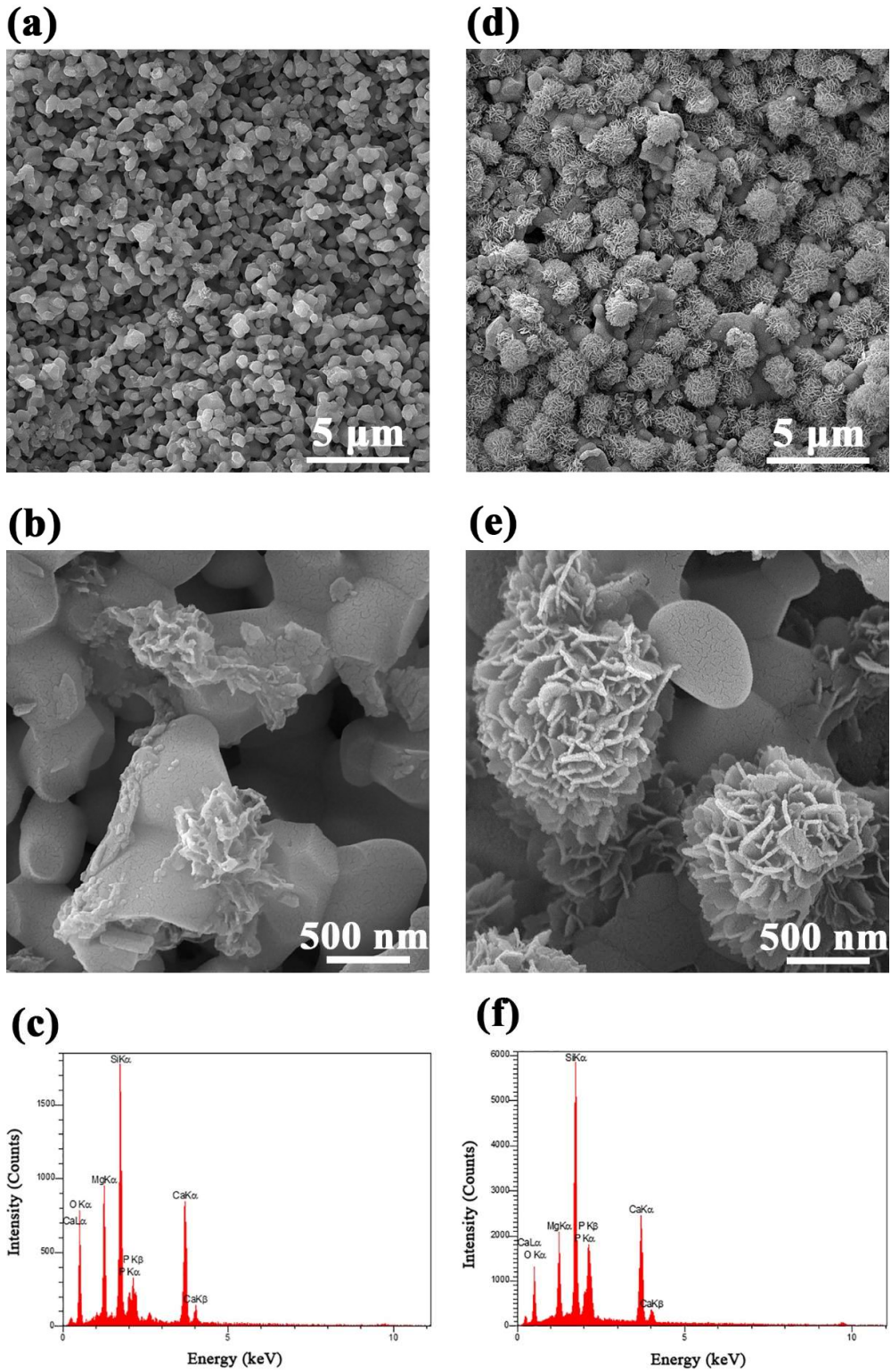

Fig. 5. SEM micrographs and EDS pattern of the sintered, undoped (a, b, c) and doped (d, e, f) samples after 3 days of immersion in SBF.

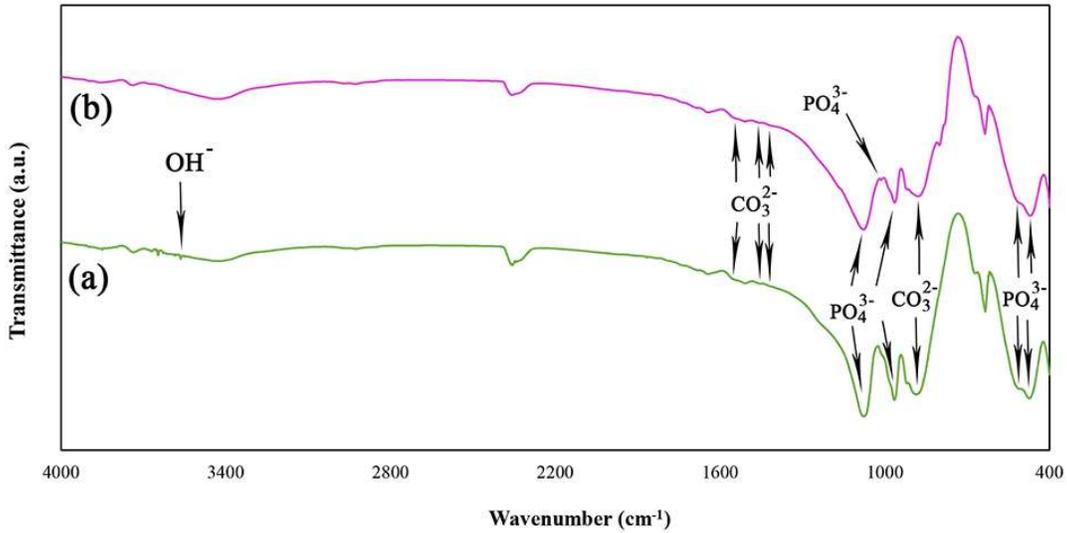

Fig. 6. FTIR spectra of the sintered, undoped (a) and doped (b) samples after immersion in the SBF.

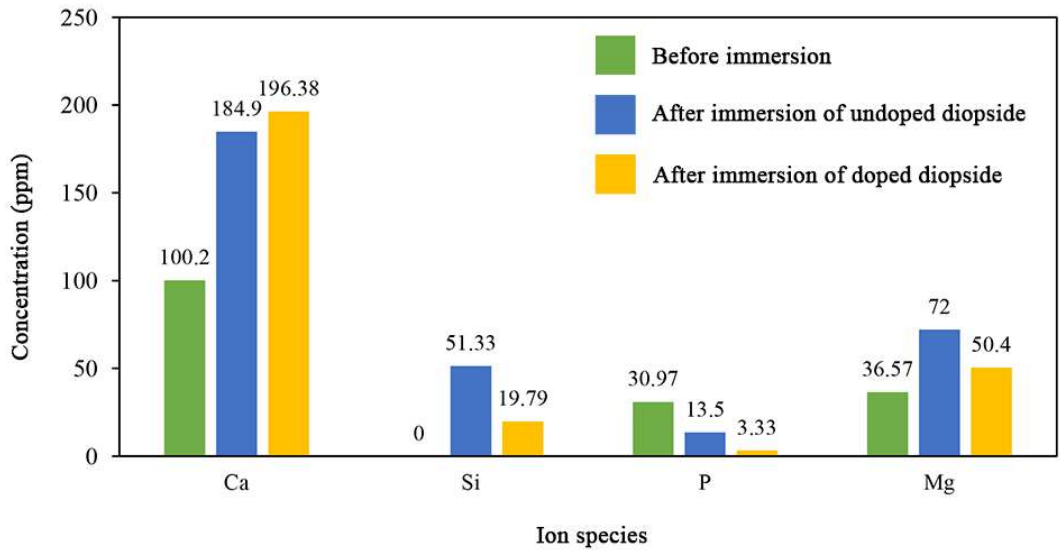

Fig. 7. ICP results on SBF after 3 days of immersion of the samples.